\documentclass[prb,twocolumn]{revtex4-1}
\usepackage[pdftex]{graphicx}
\usepackage{float}
\usepackage{amsmath}
\usepackage{amssymb}
\usepackage{amsfonts}
\usepackage{bm}
\usepackage{grffile}
\begin{document}
\author{H. Sims}
\author{S. J. Oset}
\author{W. H. Butler}
\affiliation{Center for Materials for Information Technology and Department of Physics,\\ 
University of Alabama, Tuscaloosa, Alabama 35487}
\author{James M. MacLaren}
\affiliation{Tulane University, New Orleans, Louisiana 70118}
\author{Martijn Marsman}
\affiliation{Institut f\"ur Materialphysik and Center for Computational Material Science,\\
Universit\"at Wien, Sensengasse 8, A-1090 Vienna, Austria}
\title{Determining the Anisotropic Exchange Coupling of CrO$_2$ via First-Principles Density Functional Theory Calculations}

\begin{abstract}
We report a study of the anisotropic exchange interactions in bulk CrO$_2$ calculated from first principles within density functional theory.\cite{Kohn-Sham} We determine the exchange coupling energies, using both the experimental lattice parameters and those obtained within DFT, within a modified Heisenberg model Hamiltonian in two ways. We employ a supercell method in which certain spins within a cell are rotated and the energy dependence is calculated and a spin-spiral method that modifies the periodic boundary conditions of the problem to allow for an overall rotation of the spins between unit cells. Using the results from each of these methods, we calculate the spin-wave stiffness constant $D$ from the exchange energies using the magnon dispersion relation. We employ a Monte Carlo method to determine the DFT-predicted Curie temperature from these calculated energies and compare with accepted values. Finally, we offer an evaluation of the accuracy of the DFT-based methods and suggest implications of the competing ferro- and antiferromagnetic interactions.eting ferro- and antiferromagnetic interactions.
\end{abstract}

\maketitle

\section{Introduction}\label{sec:intro}

CrO$_2$ is one of only a few known ferromagnetic oxides and is predicted to be a half-metal by first-principles calculations.\cite{DFT-CrO2} In fact, it is the only material which has been experimentally shown to be a ferromagnetic ``half-metal,''\cite{ji,gg}  a material that is a metal for one spin channel and an insulator for the other.  CrO$_2$ crystallizes in the rutile  crystal structure (Figure~\ref{rutile}), as do TiO$_2$, VO$_2$, MnO$_2$, RuO$_2$, and SnO$_2$. The existence of isostructural oxides with a variety of different electronic and magnetic properties makes the rutile system interesting for theoretical investigations of spintronics because one can envisage the growth of layered devices with the same crystal structure throughout. Since CrO$_2$ offers such special opportunities for understanding oxide spintronics, it is important to establish how well our standard electronic structure tools work in dealing with the electronic and magnetic structure of this material. It is well known that they encounter difficulties in dealing with many transition metal oxides, including the very similar oxide VO$_2$, which DFT\cite{Kohn-Sham} also predicts to be a half-metal at 0K, \cite{VO2paper} but is observed to be an insulator.  An additional motivation for understanding exchange interactions in CrO$_2$ is the fact that its Curie temperature ($T_c=386.5$ K)\cite{curie,curie2} is sufficiently close to room temperature that its magnetic properties are significantly degraded at room temperature, hindering potential spintronics applications. A better understanding may point the way to improvement.

In this work, we investigated the magnetic structure of CrO$_2$ by considering three near neighbor Cr-Cr exchange interactions: the interaction between corner and body center atoms mediated through a single oxygen atom, the interaction between a Cr and the Cr directly ``above'' it in the (001) direction, and the interaction between a Cr and its neighbor in the (100) direction. The interactions were calculated by rotating the moments of one or more of the Cr ions while constraining the others to remain parallel. We then fit the resulting energy vs.\ angle data to the Heisenberg model and extracted exchange energy parameters with a least-squares method. We also calculated the exchange interactions using a ``spin-spiral'' technique, in which a relative angular displacement was imposed upon Cr moments in adjacent cells. Similar results were obtained with both approaches.  The calculated $T=0$ K exchange interactions were subsequently used to determine the magnetization as a function of temperature via low-T spin-wave dispersion and a Monte-Carlo method.

\section{Electronic Structure of C\lowercase{r}O$_2$ Within Density Functional Theory}\label{sec:estruct}

\begin{figure}[ht]
\includegraphics[]{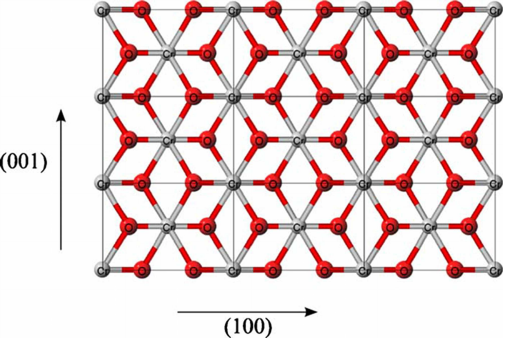}
\caption[]{Rutile structure projected onto the x-z plane. For CrO$_2$, we use $a=4.42$ and $\frac{c}{a}\approx0.670$ (experimental parameters). The oxygen octahedra can be clearly seen surrounding each Cr ion. The terms ``corner'' and ``body-center,'' used throughout this work, refer to the Cr ions at the corner and center of the rectangular cells seen here.}
\label{rutile}
\end{figure}

In the following, the electronic structure and density of states of CrO$_2$ were calculated using density functional theory\cite{Kohn-Sham} (DFT) and the generalized gradient approximation\cite{Perdew-Wang} (GGA) using GGA-relaxed lattice parameters (see Table\ \ref{tab:everything}).  Our calculated density of states is similar to previous calculations.\cite{DFT-CrO2,inorg}  For a detailed discussion of the electronic structure of the rutiles, we refer the reader to the work of Sorantin and Schwarz.\cite{inorg} Additionally, the lattice structure is presented in Figure~\ref{rutile}.

\begin{figure}[ht]
\includegraphics[]{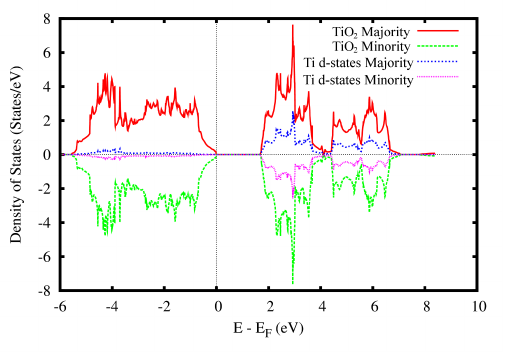}
\caption[]{Density of States for rutile TiO$_2$ calculated within DFT using the GGA (with GGA-relaxed lattice parameters).}
\label{DOSTiO2}
\end{figure}

It is straightforward to show that if we treat this system in a tight-binding approximation in which the TM atoms only interact directly with the oxygen atoms (i.e.~hopping matrix elements only connect nearest neighbors), there will be an energy gap separating the oxygen $p$-states and the TM $d$-states. The gap extends from the O-$p$ onsite energy to the TM-$d$ onsite energy.  This gap is apparent in TiO$_2$, for which the oxygen $p$-states are filled and the Ti $d$-states are empty (Fig.\ \ref{DOSTiO2}).  When an energy gap occurs at the Fermi energy, it contributes significantly to reducing the energy of the structure, because all occupied states are pushed down in energy, while all unoccupied states are pushed up.  In CrO$_2$, there are two additional electrons per TM atom compared to TiO$_2$, so some of the $d$-states above the gap must be occupied.

\begin{figure}[ht]
\includegraphics[]{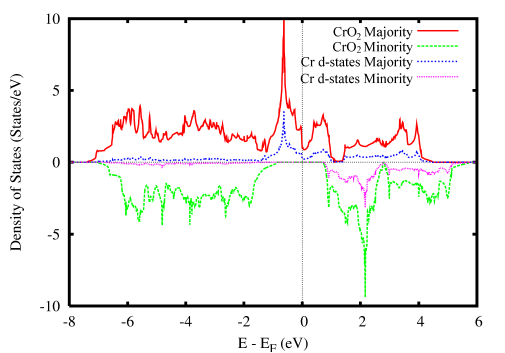}
\caption[]{Density of States for Ferromagnetic CrO$_2$ calculated within DFT using the GGA (with GGA-relaxed lattice parameters).}
\label{DOSFMCrO2}
\end{figure}

Comparing these the energies of the possible magnetic configurations (FM, AF, or nonmagnetic) using total-energy GGA DFT calculations (with GGA-relaxed lattice parameters), it is not surprising that we find that the ferromagnetic state has the lowest energy (with the DOS seen in Figure~\ref{DOSFMCrO2}), the nonmagnetic state the highest (1.02 eV above ferromagnetic) with the anti-ferromagnetic intermediate between the two (0.30 eV above ferromagnetic). Thus, the tendency to form a moment in CrO$_2$ is very strong, and the energy associated with the ferromagnetic alignment of moments based on this initial test is moderately large within DFT.  It should be recognized that other more complicated spin arrangements (e.g. different antiferromagnetic states) may have lower energy than the simple one calculated here.

\section{Exchange Interactions in C\lowercase{r}O$_2$}\label{sec:exch}

In order to investigate interatomic exchange interactions in CrO$_2$ in more detail, we have calculated the near-neighbor exchange interactions along the (100), (001), and (111) directions by rotating moments within specially-constructed supercells. We fit the resulting relationship between the energy of the system and the angle of rotation to the Heisenberg model
\begin{equation}
H = -\sum_{i,j}J_{ij}\boldsymbol{\mu}_i \cdot \boldsymbol{\mu}_j
\label{eq:heis}
\end{equation}
where $\left|\boldsymbol{\mu}\right| = g\mu_B S = 2\mu_B$ is the spin moment, $g$ is the electron spin $g$-factor, $S = 1$ is the spin number, and $\mu_B$ is the Bohr magneton. To make contact with the standard Heisenberg model, we can pull the magnitude of the spin moment ($2\mu_B$) into the value of $J$ and treat the spins as unit vectors.

In addition to this supercell approach, we have taken advantage of a recently developed feature in the Vienna {\em Ab-initio} Simulation Package\cite{VASP} (VASP) to calculate a so-called helimagnetic state in which the moment in the $n^{\mathrm{th}}$ magnetic layer is canted by an angle $n\phi$ with respect to the $0^{\mathrm{th}}$ layer. In so doing, we are able to calculate several orders of $J_n$ of the form
\begin{equation}
E = E_0 + \sum_nJ_n\cos{}n\phi
\end{equation}
via Fourier analysis. The relationship between the $J_n$ and the $J_{ij}$ will be made explicit in Section~\ref{ssec:heli}.

\subsection{Near Neighbor Exchange Using Supercells}\label{ssec:supercell}
\begin{figure}[ht]
\includegraphics[]{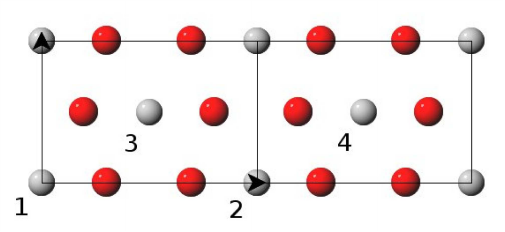}
\caption{The (100) supercell projected onto the x-z plane, with Cr ions numbered for comparison to Table\ \ref{exchtab}.}\label{100fig}
\end{figure}

\begin{figure}[ht]
\includegraphics[]{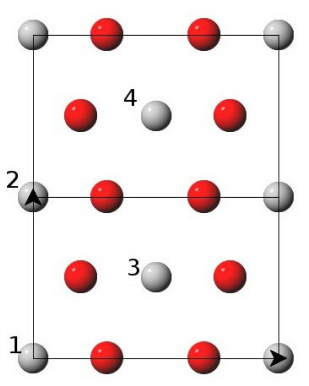}
\caption{The (001) supercell projected onto the x-z plane, with Cr ions numbered for reference.}\label{001fig}
\end{figure}

All of the calculations in this study were performed within DFT\cite{Kohn-Sham} in the GGA\cite{Perdew-Wang} and in the local (spin) density approximation with onsite Coulomb interactions (LSDA+U)\cite{LDApU} using the Dudarev method,\cite{theDude} for which we use $U-J$ = 2.1 eV, in agreement with the $U$ and $J$ values seen in other works.\cite{UJ} We perform all calculations using the VASP software\cite{VASP} and pseudopotentials generated by Kresse {\emph et al}.\cite{Kresse}. To calculate the near-neighbor exchange interactions, we created a supercell containing two rutile unit cells (using both experimental and DFT-relaxed lattice parameters), stacked in either the (100) (Fig.\ \ref{100fig}) or (001) (Fig.\ \ref{001fig}) direction as appropriate. In all of the following calculations, we use an energy cut-off of 500 eV. For cells stacked along the (100) direction, we use a $5\times9\times15$ Monkhorst-Pack\cite{MP} grid of k-points, a $9\times9\times7$ grid for supercells stacked along (001), and a $9\times9\times15$ grid for the 6-atom cell used in the spin-spiral calculations. We also make use of the spin interpolation method of Vosko-Wilk-Nusair.\cite{VoskoWN} Each of the 12-atom supercells has four Cr ions, whose magnetic moments we can individually constrain within the calculation. We chose three distinct magnetic configurations designed to probe the exchange coefficients. In the first configuration, we rotated the moment of a corner Cr atom and held all other moments fixed using the constraining field method in VASP. In the second, we rotated the two Cr moments in the centers of their respective unit cells, and in the third we rotated a corner atom and its nearest center atom. A summary of the configurations can be found in Table~\ref{exchtab}. 

\begin{table}[ht]  
\begin{tabular}{lcccc}
\hline
 & {\bf Cr$_1$} & {\bf Cr$_2$} & {\bf Cr$_3$} & {\bf Cr$_4$} \\
\hline
{\bf Case 1} & fixed & rotated & fixed & fixed \\
{\bf Case 2} & rotated & rotated & fixed & fixed \\
{\bf Case 3} & rotated & fixed & rotated & fixed \\
\hline
\end{tabular}
\caption[]{Magnetic configurations used to calculated exchange coupling. The numbers are as indicated in Figures~\ref{100fig} and \ref{001fig}.}\label{exchtab}
\end{table}

\begin{figure}[ht]
\includegraphics[]{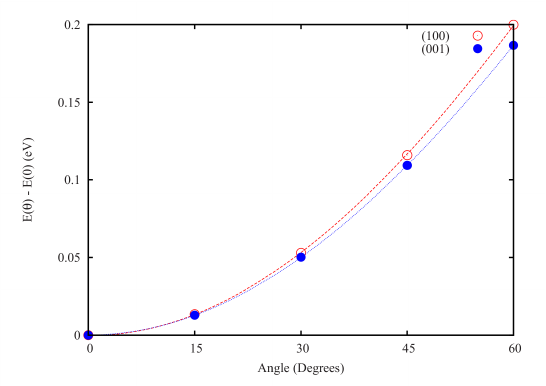}
\caption{Energy vs. angle between rotated and fixed moments for Case 1. The curve is the fit to $A(1-\cos\theta)+B$.}\label{fig:case1}
\end{figure}

\begin{figure}[ht]
\includegraphics[]{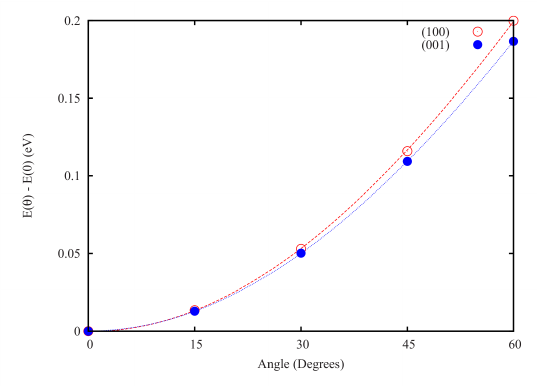}
\caption{Energy vs. angle between rotated and fixed moments for Case 2. The curve is the fit to $A(1-\cos\theta)+B$.}\label{fig:case2}
\end{figure}

\begin{figure}[ht]
\includegraphics[]{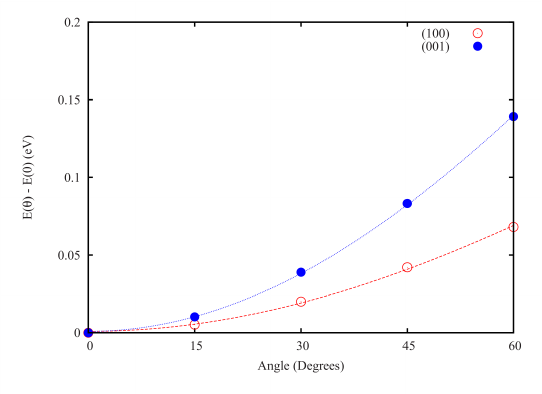}
\caption{Energy vs. angle between rotated and fixed moments for Case 3. The curve is the fit to $A(1-\cos\theta)+B$.}\label{fig:case3}
\end{figure}

To ensure that we can accurately apply our modified Heisenberg model to these systems, we rotated the moments through small angles (up to 60$^\circ$). We fit the energy vs.\ angle data to $A(1-\cos\theta)+B$, where $A$ is the contribution to the exchange energy from all rotated moments and $B$ is simply the angle-independent component of the energy. The fits can be seen in Figures \ref{fig:case1} - \ref{fig:case3}.

For a given choice of supercell orientation, we have the following system of equations:
\begin{equation}
A_{\text{Case 1}}=8J_{111}+2J_{100/001}
\label{eq:100fit}
\end{equation}
\begin{equation}
A_{\text{Case 2}}=16J_{111}
\end{equation}
\begin{equation}
A_{\text{Case 3}}=8J_{111} + 4J_{100/001}
\end{equation}
Using a least-squares technique for overdetermined systems of equations,\cite{Williams} we can write
\begin{equation}
AJ=b
\label{eq:lsquare}
\end{equation}
\begin{equation}
A^TAJ=A^Tb
\end{equation}
\begin{equation}
\overline{J}=(A^TA)^{-1}A^Tb
\end{equation}
\begin{equation}
\sigma = \left|A\overline{J}-b\right|
\end{equation}
where $\overline{J}$ is the calculated $J$ column vector, $\sigma$ is the error in the fit, and
\begin{equation}
A=
\left(\begin{array}{cc}
8 & 2 \\
16 & 0 \\
8 & 4
\end{array}\right)\qquad
J=
\left(\begin{array}{c}
J_{111} \\
J_{100/001}
\end{array}\right)
\end{equation}

We summarize the calculations performed within GGA and LSDA+U for experimental and relaxed lattice parameters using the supercell method in Table~\ref{tab:everything}. Throughout this work, the terms ``experimental'' and ``relaxed'' (in the sense used in Table \ref{tab:everything}) denote structures with the experimental and the GGA- or LSDA+U-relaxed lattice parameters, respectively.

\begin{table*}[ht]
\begin{tabular}{lcccc}
\hline
 & \multicolumn{2}{c}{GGA} & \multicolumn{2}{c}{LSDA+U} ($U-J = 2.1$ eV) \\
 &  Experimental & Relaxed & Experimental & Relaxed \\
a (\AA) & 4.421 & 4.4495 & 4.421 & 4.3775 \\
c (\AA) & 2.917 & 2.9470 & 2.917 & 2.8758 \\
$J_{100}$ (meV) & $-11.8\pm2.5$ & $-10.4\pm0.7$ & $-2.0\pm1.0$  & $-2.4\pm0.8$ \\
$J_{001}$ (meV) & $33.8\pm5.6$ & $33.8\pm5.0$ & $35.6\pm1.5$  & $33.1\pm1.0$ \\
$J_{111}$ (meV) & $23.2\pm6.1$ & $22.9\pm5.0$ & $24.2\pm1.5$  & $24.4\pm1.0$ \\
\hline
\end{tabular}
\caption[]{Definition of ``experimental'' and ``relaxed'' lattice parameters and summary of all calculated exchange energies obtained using the supercell method. Uncertainties given arise from the error in the least-squares fit. Additionally, in $J_{111}$, there is some (usually negligible) contribution to the error from the standard deviation of the values obtained through (100)- and (001)-stacked supercells. Note that the (100) and (010) directions are equivalent and are referred to as (100) throughout this work.}\label{tab:everything}
\end{table*}

The results of the calculations for the three cases are summarized as follows: in each case, we find a near-perfect fit to the cosine function, provided that we restrict the fit to small angles (less than or equal to $60^\circ$), as we did with the original calculations. We can see the anisotropic nature of the exchange clearly in Table~\ref{tab:everything}, which is to be expected given the shape of the cell. Most interestingly, we find that the interaction between Cr neighbors along the (100) or (010) directions (parallel to the $a$ or $b$ axes) is antiferromagnetic. However, the strength and multiplicity of the other interactions is sufficient to lead to a ferromagnetic ground state. Considering the dependence on lattice parameter, we notice that the (001) and (111) interactions seem to be almost unchanged with the small (0.6\%) change in lattice constant. Somewhat surprisingly, however, the (100) interaction (calculated within the GGA) increases (becomes more positive) by more than an meV under this small expansion of the lattice. We also note that the LSDA+U calculations predict a smaller (in magnitude), though still negative, $J_{100}$.

\subsection{Helimagnetism}\label{ssec:heli}

Helimagnetism is a noncollinear magnetic state in which the spins in adjacent layers along a certain direction are rotated with respect to one another by a fixed angle. Rutile MnO$_2$, for example, has been shown to exhibit helimagnetic ordering in the ground state.\cite{MnO2} We do not suspect that CrO$_2$ is a helimagnetic material, but by setting up a helimagnetic spin state, we can investigate the exchange using a different approach. The recently-added spin spiral capabilities of VASP\cite{spiral} allow us to calculate arbitrarily long-range exchange interactions within bulk CrO$_2$.

The spin spiral method modifies the periodic boundary conditions of the supercell approach, imposing helimagnetic order on the magnetic structure as determined by the propagation vector {\bf q}. The vector {\bf q} and the angle $\phi$ between any two spins are given by
\begin{equation}
\phi=\mathbf{q}\cdot\mathbf{r}_{j}
\end{equation}
\begin{equation}
\mathbf{q}=\frac{2\pi}{a_i}\xi\hat{\mathbf{e}}_i
\label{qeq}
\end{equation}
where the polar angle $\theta$ is restricted to $\frac{\pi}{2}$ ($\mu_z=0$). Thus, the moment of an ion is given by
\begin{equation}
\boldsymbol{\mu}_{\boldsymbol{r}_i}(\boldsymbol{q}) = \hat{\mathbf{e}}_x \mu\cos(\boldsymbol{q}\cdot\boldsymbol{r}_i) + \hat{\mathbf{e}}_y \mu\sin(\boldsymbol{q}\cdot\boldsymbol{r}_i)
\label{eq:mom}
\end{equation}
where $\mu = 2 \mu_B$ and $r_0 = 0$.

In defining $\mathbf{q}$, we choose the unit vector $\hat{\mathbf{e}}_i$ to be either the (100) or (001) direction, and allow $\xi$ to vary between 0 and 1. Clearly, when $\xi = 0$, we recover the ferromagnetic state.

\begin{figure}[ht]
\includegraphics[]{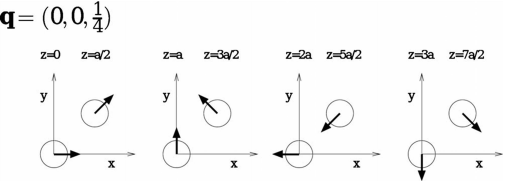}
\caption[]{A schematic representation of a spin spiral setup. The left-most cell is all that is needed for the calculation; the others merely illustrate the propagation of the spiral throughout the lattice.}\label{fig:spiralexample}
\end{figure}

Because the unit cell contains two magnetic ions, varying the angles between neighboring CrO$_2$ cells requires that one modify both $\xi$ and the orientation of the magnetic moments in the 0$^{\text{th}}$ cell. For example, to obtain a system in which neighboring magnetic ``layers'' (one half of a unit cell) are oriented at an angle of $\frac{\pi}{4}$ from one another, we use $\xi=\frac{1}{4}$, so that each cell after the initial one is rotated by $\frac{\pi}{2}$. We then set up the moments in the initial cell such that the corner and body-centered Cr moments are oriented at the desired angle of $\frac{\pi}{4}$, leading to a smooth spin wave in the desired direction. This can be seen schematically in Figure~\ref{fig:spiralexample}.

In this work, we choose a relatively short spin wavelength in order to simplify the analysis, although the method allows for more general configurations as well. Using different values of {\bf q}, and thus different values of $\theta$, we create a {\bf q} spectrum. We then use Fourier analysis to extract the $J_n$. These $J_n$ differ in meaning from the $J$s calculated using the supercell method; they are given by
\begin{equation}
J_1 = 8 J_{111}
\end{equation}
\begin{equation}
J_2 = 2 J_{100/001}
\end{equation}
To calculate the helimagnetic state, we used a supercell composed of a single rutile unit cell. The angle of each subsequent Cr ion with respect to the first is given by \eqref{eq:mom}. After acquiring $N=5$ points (including the zero-frequency point q=0) of the $E(q)$ curve, we performed a discrete Fourier transform to obtain the first 4 $J_n$. We used a discrete cosine transform of the first kind (appropriate when the data are even about the end-points), given by
\begin{equation}
J_n=\frac{1}{4}\left(E_0 + (-1)^nE_{N-1}\right) + \frac{1}{2}\sum_{j=1}^{N-2}E_j\cos\left(\frac{\pi}{N-1}jn\right)
\end{equation}
where the $J_n$ are the exchange energies and the $E_i$ are the calculated $E(\mathbf{q}_i)$.

We find good agreement between the $J_1$ calculated with (100) and (001) spin spirals, as expected. We also find a difference in sign between $J_2$ in the (100) and (001) cases, in agreement with the larger supercell calculations. Moreover, this method yields the additional parameters $J_3$ and $J_4$, corresponding to $8J_{211/112}$ and $2J_{200/002}$, respectively. These higher-order energies are smaller than the first- and second-order exchange energies, and will be neglected in further analysis. The results of the calculations are summarized in Table~\ref{tab:heli}. 

\begin{table}[ht]
\begin{tabular}{lcccc}
\hline
 & \multicolumn{2}{c}{GGA} & \multicolumn{2}{c}{LSDA+U} \\
 & Experimental & Relaxed & Experimental & Relaxed \\
\hline
{\bf $J_{100}$} (meV) & -12.0 & -12.2 & -6.9 & -6.8 \\
{\bf $J_{001}$} (meV) & 27.5 & 29.8 & 32.6  & 28.4  \\
{\bf $J_{111}$} (meV) & 20.8 & 20.7 & 26.0 & 25.9 \\
\hline
\end{tabular}
\caption{Summary of calculated exchange interactions (in meV) using the spin spiral method (compare with Table~\ref{tab:everything}). Errors in these numbers would arise from errors in the VASP total energy calculations, which are on the order of 1 meV. Note that the effect of the change in lattice parameter is smaller in the spin spiral method. Using GGA, the spin-spiral $J_{100}$, $J_{001}$, and $J_{111}$ fall inside or nearly inside the error bars for the super cell calculations. In LSDA+U, however, the $J_{100}$ are about three times larger (more negative).}\label{tab:heli}
\end{table}

\section{Comparison with Experiments}\label{sec:exptcomp}

\subsection{Spin Wave Stiffness}\label{ssec:sw}

To compare our calculations against known experimental results, we have calculated the spin wave stiffness constant for CrO$_2$ using expressions similar to those derived by Schlottmann\cite{schlott}:
\begin{equation}
D_{100} = 2(J_{111} + J_{100}) S a^2 \label{eq:100D}
\end{equation}
\begin{equation}
D_{001} = 2(J_{111} + J_{001}) S c^2\label{eq:001D}
\end{equation}
where $a$ and $c$ are the lattice spacings in the appropriate directions and $S$ is the spin number (1 for CrO$_2$). These expressions can be easily understood as anisotropic extensions of results obtained for magnons in a one-dimensional chain (for which $D=2JSa^2$). In his work, Schlottmann considers the spins as quantum operators, and he keeps the value of $\mathbf{S}_i\cdot\mathbf{S}_j$ separate from $J$. Additionally, he neglects $J_{100}$ in his expression for $D_{100}$. However, we use classical spins of magnitude $2 \mu_B$ (although the units are collapsed into the exchange constant $J$ as previously explained). Consequently, we must scale our $J$s by $1/\left|\boldsymbol{\mu}\right|^2 = 1/4$ in order to apply this expression. Further, our calculations indicate that $J_{100}$ is not negligible when compared to $J_{111}$ and $J_{001}$, so we have included it in our analysis. Using this model, we calculate $D_{100}$ and $D_{001}$ for the various cells, exchange-correlation approximations, and methods considered throughout this work. Table~\ref{tab:stiff} reviews the values we obtained. Examining the experimental literature, we find several values (in good agreement with one another) for the spin wave stiffness obtained through different methods. All of the experimental values assume an isotropic stiffness constant. Ji \textit{et al.}\cite{stiff} fit the M(T) curve in order to obtain the coefficient on the $T^{3/2}$ term, from which they determine $D=1.8\times10^{-40}$ Jm$^2$. Zou \emph{et al}\cite{zou} used magnetic force microscopy to determine the length and width of domain walls in CrO$_2$, from which they were able to calculate $D = 2.62\times10^{-40}$ Jm$^2$. Further, Rameev \emph{et al.}\cite{fmr} used ferromagnetic resonance to measure the bulk magnon modes and obtained $D_B=3\times10^{-10}$ Oe cm$^2$, which is equivalent to $D=0.57\times10^{-40}$ Jm$^2$ via the relation $D_B = 2A / \mu_0 M_s$,\cite{units} which is smaller than but of the same order as the other reported values.

\begin{table*}[ht]
\begin{tabular}{lccccc}
\hline
 & & \multicolumn{2}{c}{GGA} & \multicolumn{2}{c}{LSDA+U} \\
 & & Experimental & Relaxed & Experimental & Relaxed \\
\hline
{\bf Supercell} & $\bm{D_{100}}$ $(\times10^{-40}$ J m$^2)$ & $1.81$ & $1.96$ & $3.48$ & $3.38$ \\
 & $\bm{D_{001}}$ $(\times10^{-40}$ J m$^2)$ & $3.91$ & $3.87$ & $4.08$ & $3.81$ \\
 & $\bm{D_{avg}}$ $(\times10^{-40}$ J m$^2)$ & $2.34$ & $2.46$ & $3.67$ & $3.52$ \\ 
{\bf Spin Spiral} & $\bm{D_{100}}$ $(\times10^{-40}$ J m$^2)$ & $1.38$ & $1.35$ & $3.00$ & $2.94$ \\
 & $\bm{D_{001}}$ $(\times10^{-40}$ J m$^2)$ & $3.29$ & $3.46$ & $3.99$ & $3.67$ \\
 & $\bm{D_{avg}}$ $(\times10^{-40}$ J m$^2)$ & $1.84$ & $1.84$ & $3.30$ & $3.17$ \\
\hline
\end{tabular}
\caption{Comparison of the calculated spin stiffness constants $D$ for different methods of first-principles calculation. Here, $D_{avg} = \left(D_{100}\sqrt{D_{001}}\right)^{2/3}$.}\label{tab:stiff}
\end{table*}

\begin{figure}[ht]
\includegraphics[]{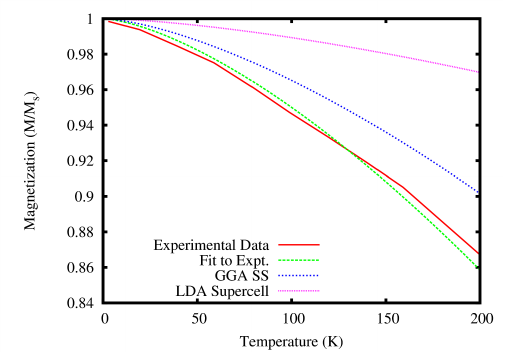}
\caption[]{The low-temperature M(T) curve. The GGA Spin Spiral and LSDA+U Supercell curves represent the extremes of the range of calculated M(T) curves. We compare against the actual experimental data\cite{curie} and a low-T fit to these data.}\label{fig:magnon}
\end{figure}

Using our calculated $D$s, we can predict the low-temperature spin-wave contribution to the magnetization as a function of temperature. The relatively straight-forward generalization of the argument found in Kittel\cite{Kittel} for a cubic system that we used above to calculate the spin stiffness also allows one to write the spin-wave dispersion relation for small excitations and long wavelengths as
\begin{align}
\omega(k,k_z) = D k^2 + D_z k_{z}^{2}\\
k^2 \equiv k_{x}^{2} + k_{y}^{2}\\
D \equiv D_{100}\\
D_z \equiv D_{001}
\label{eq:disp}
\end{align}
Integrating over a surface of constant $\omega$ in $k$-space, one obtains a density of states given by
\begin{equation}
N(\omega) = \frac{1}{4\pi^2}\frac{1}{D\sqrt{D_z}}\sqrt{\omega}
\end{equation}
Using this expression and the Planck distribution, we can calculate the coefficient $B$ in the $T^{3/2}$ model
\begin{equation}
M(T) = M(0)(1-BT^{3/2})
\end{equation}
\begin{widetext}
\begin{equation}
B=\frac{0.0587}{SQ}\frac{1}{2S(J_{100}+J_{111})}\frac{1}{\sqrt{2S(J_{001}+J_{111})}}k_{B}^{3/2} = \frac{0.0587}{SQ}\frac{V}{D\sqrt{D_z}}k_{B}^{3/2}
\end{equation}
\end{widetext}
where $Q$ is the number of magnetic ions per unit cell (2, in this case), $V$ is the volume of the cell, and $k_B$ is Boltzmann's constant. Fitting the experimental\cite{stiff} M(T) curve yields $B=5\times10^{-5}$ K$^{-3/2}$. Using the spin-wave stiffnesses shown in Table~\ref{tab:stiff}, we have, for supercells, $B_{GGA}^{expt}=2.40\times10^{-5} $K$^{-3/2}$, $B_{GGA}^{rel}=2.27\times10^{-5}$ K$^{-3/2}$, $B_{LSDA+U}^{expt}=1.22\times10^{-5}$ K$^{-3/2}$, and $B_{LSDA+U}^{rel}=1.26\times10^{-5}$ K$^{-3/2}$. For the spin spiral approach, $B_{GGA}^{expt}=3.43\times10^{-5}$ K$^{-3/2}$, $B_{GGA}^{rel}=3.48\times10^{-5}$ K$^{-3/2}$, $B_{LSDA+U}^{expt}=1.43\times10^{-5}$ K$^{-3/2}$, and $B_{LSDA+U}^{rel}=1.47\times10^{-5}$ K$^{-3/2}$. Thus, the coefficient obtained from GGA is within a factor of two, while that derived from LSDA+U is off by about a factor of four. Assuming that DFT overestimates each exchange energy equally, this implies that our calculated values of $J$ may differ from experimental values by about 50\% for GGA and a factor of about $2.5$ for LSDA+U (with $U-J=2.1$ eV). In each case, the spin spiral numbers are closer to experiment. Figure~\ref{fig:magnon} shows the low-T M(T) curves from the calculated spin-wave dispersion compared to that from a fit to the experimental M(T) curve.

\subsection{Curie Temperature}\label{ssec:curie}

In light of the favorable agreement between calculated and experimental spin stiffness, we subsequently attempted to calculate the magnetic ordering temperature of CrO$_2$, comparing a mean field prediction to Monte Carlo simulations. A mean-field model using the calculated exchange parameters yields a Curie temperature several times larger than the measured value of 386.5 K.\cite{curie,curie2} The mean-field expression is given by
\begin{equation}
k_BT = \frac{3}{2}J_{tot}
\end{equation}
where $J_{tot}$ is equivalent to half of the energy difference between a ferro- and an antiferromagnetic configuration in a 6-atom (2-Cr) cell. Using this expression, we obtain a mean-field Curie temperature for CrO$_2$ of 1160 K or 1240 K for the experimental and DFT-relaxed lattice parameters in the supercell method, respectively. This is somewhat surprising given the above analysis of our estimation of the exchange. However, it is not sufficient to consider only the low-temperature behavior. In order to gain a simple yet illuminating picture of the temperature dependence, we utilized a Monte Carlo simulation using the Metropolis-Hastings algorithm\cite{metropolis} with random numbers generated using the Mersenne Twister method.\cite{mersenne} For this simulation, we used a cubic grid of $10\times10\times10$ unit cells (L=10), where a unit cell consists of a corner and body-centered Cr ion.  Only Cr ions are considered, and they are treated as simple constant-magnitude magnetic moments. Our first-principles calculations indicate that the constant-magnitude approximation should be valid as long as the angle between adjacent moments is less than $100^\circ$.

We begin with a random spin configuration with the spin vectors chosen to be uniformly distributed on the unit sphere. In the Metropolis method, an iteration consists of a randomly-chosen Cr ion being assigned a magnetic moment in a random direction. This will result in a change in energy $\Delta$E from the old configuration. If $\Delta$E is negative, meaning the new energy is lower, the new direction for that moment is kept. Otherwise, the new direction still has a probability of $e^{-\Delta{}E/k_BT}$ of being kept in its new orientation to simulate thermal agitation. If neither condition for keeping the moment's new direction is met, then the change is undone, and the lattice of spins remains unmodified until the next iteration. Following a ``burn-in'' period to remove any artifacts of the initial configuration, we take averages of the magnetization at regular intervals to allow for the computation of thermodynamic quantities.

\begin{figure}[ht]
\includegraphics[]{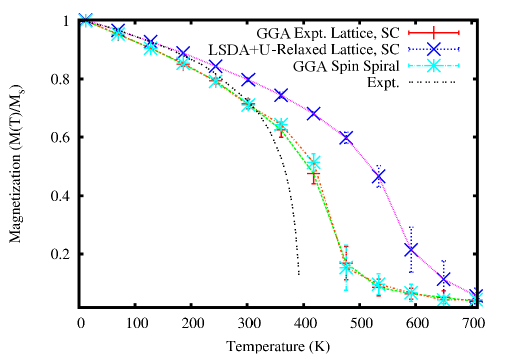}
\caption{The calculated M(T) behavior using the Monte Carlo method. We present supercell (SC) results using experimental lattice parameters within GGA, spin spiral results using GGA-relaxed lattice parameters, and supercell results using LSDA+U-relaxed lattice parameters to indicate the range of results obtained. We also compare these data with experiment\cite{curie}.}\label{heiscurie}
\end{figure}

\begin{figure}[ht]
\includegraphics[]{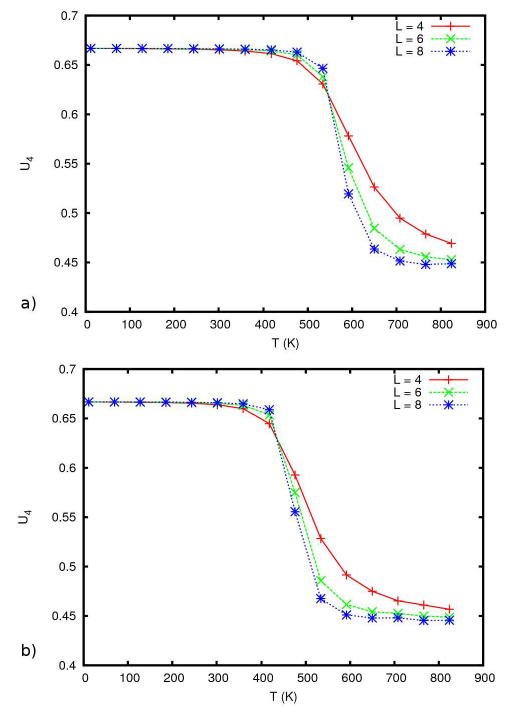}
\caption{Plots of the Binder cumulant vs.~ simulation temperature for (a) supercell (b) spin spiral calculations for L = 4, 6, and 8. Both plots were obtained using the exchange coupling values for the GGA-relaxed lattice. The point of intersection of the three curves gives the true Curie temperature for the simulation.}\label{fig:u4}
\end{figure}

\begin{figure}[ht]
\includegraphics[]{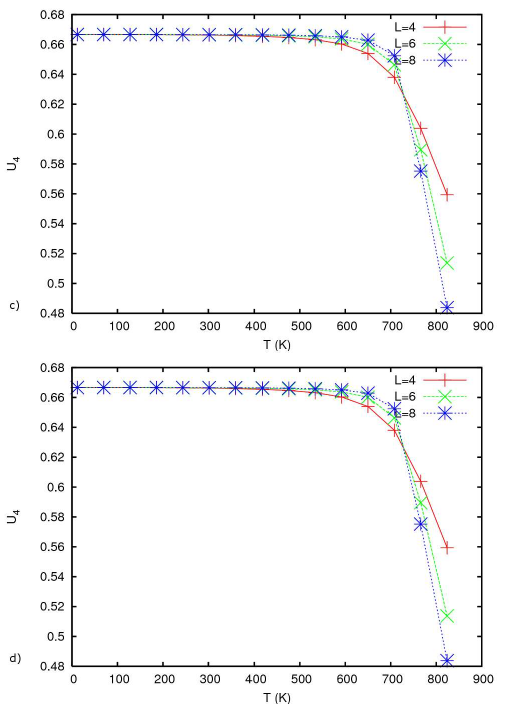}
\caption{Plots of the Binder cumulant vs.~ simulation temperature for (c) supercell (d) spin spiral calculations for L = 4, 6, and 8 using LSDA+U-relaxed lattice parameters.}\label{fig:ldau4}
\end{figure}

The calculation of $\Delta$E at each step considers all nearest neighbors along (100), (010), (001), and (111) directions, using a Heisenberg interaction between moments with the calculated exchange constants for GGA and LSDA+U with experimental and DFT-relaxed lattice parameters. Figure~\ref{heiscurie} shows the simulated results for the magnitude of the net magnetization versus temperature compared to reported values.\cite{curie} When interpreting these data, one must must be cognizant of the fact that the Monte Carlo simulations exhibit several shortcomings---namely, that it will necessarily not be able to predict the correct low-temperature T-dependence (as it uses a classical model), that there exists an unphysical tail on the curve arising from finite-size effects in the lattice, and that we assume that exchange remains constant with temperature, likely leading to an overestimation of the Curie temperature. The errors in the shape of the curve at low temperature should not have an impact in the accuracy of the result, as each value of $k_BT$ is run independently. Further, the high-temperature tail can be accounted for by calculating the Binder cumulant\cite{Binder} instead of the raw magnetization. The Binder cumulant is given by
\begin{equation}
U_4 = 1-\frac{\langle{}m^4\rangle}{3\langle{}m^2\rangle^2}
\end{equation}
By calculating $U_4$ as a function of temperature for a range of L, we can find the true calculated critical temperature at the intersection of the resulting curves (Figures \ref{fig:u4} and \ref{fig:ldau4}). The remaining discrepancy, that the exchange will reduce in strength as temperature rises, is a limitation of exploring this behavior with first-principles calculations.

\section{Conclusions}\label{sec:con}

We have calculated the near neighbor exchange interactions for bulk CrO$_2$ in the (100), (001), and (111) directions. From our calculated spin stiffness parameters and the results of our classical Heisenberg Metropolis method, we obtain some confidence that DFT and VASP can describe the exchange coupling in CrO$_2$ (to within 15\% using the GGA-spin-spiral method). However, the agreement is not in all cases impressive (for example in the LSDA+U calculations). One should understand that, although DFT is well-suited to determine the structural parameters of such a system (less than 1\% error in the determination of the lattice parameters), it is known to underestimate band gaps (such as that in the minority channel of CrO$_2$), and it is possible that the exchange coupling (particularly the double exchange between Cr-O-Cr neighbors) may arise from correlation effects that DFT is ill-suited to handle. Given such considerations, an error as low as 15\% (in one case) could be considered a modest success.

Examining the calculated exchange parameters, we find that the sign of $J_{100}$, both in the supercell and the equivalent spin spiral calculations, indicates the possibility of non-collinear behavior in CrO$_2$ if the exchange parameters are modified. Thus, a mixed interface between CrO$_2$ and another material (such as RuO$_2$) might lead to non-collinear spins if the ratio between nearest and next-nearest neighbor interactions is pushed into a ``favorable'' zone. We investigate this possibility explicitly for CrO$_2$-RuO$_2$ interfaces in an upcoming paper. Non-collinear spins in the neighborhood of a spacer material would eliminate the expected GMR effect in such a system.

\section{Acknowledgments}
This work was supported by the NSF-DMR under Grant No.\ 0706280 and MRSEC Grant No.\ 0213985. It was completed using computing resources from the University of Alabama's High Performance Cluster.
\newline

\end{document}